\documentclass[reprint,groupedaddress,superscriptaddress,prx]{revtex4-1}

\usepackage{amsmath}  \DeclareMathOperator\arctanh{arctanh}
\usepackage{braket}
\usepackage{graphicx}
\usepackage{units}
\usepackage[colorlinks=true,citecolor=blue,urlcolor=blue,linkcolor=blue]{hyperref}

\begin{document}

\title{Experimental test of exchange fluctuation relations in an open quantum system}

\author{S. Hern\'{a}ndez-G\'{o}mez}
\affiliation{European Laboratory for Non-linear Spectroscopy (LENS), and Department of Physics and Astronomy, Universit\`a di Firenze, I-50019 Sesto Fiorentino, Italy}
\affiliation{Istituto Nazionale di Ottica del Consiglio Nazionale delle Ricerche (CNR-INO), I-50019 Sesto Fiorentino, Italy}
\author{S. Gherardini}
\affiliation{European Laboratory for Non-linear Spectroscopy (LENS), and Department of Physics and Astronomy, Universit\`a di Firenze, I-50019 Sesto Fiorentino, Italy}
\affiliation{Scuola Internazionale Superiore di Studi Avanzati (SISSA), I-34136 Trieste, Italy}
\author{F. Poggiali}
\affiliation{European Laboratory for Non-linear Spectroscopy (LENS), and Department of Physics and Astronomy, Universit\`a di Firenze, I-50019 Sesto Fiorentino, Italy}
\affiliation{Istituto Nazionale di Ottica del Consiglio Nazionale delle Ricerche (CNR-INO), I-50019 Sesto Fiorentino, Italy}
\author{F. S. Cataliotti}
\affiliation{European Laboratory for Non-linear Spectroscopy (LENS), and Department of Physics and Astronomy, Universit\`a di Firenze, I-50019 Sesto Fiorentino, Italy}
\affiliation{Istituto Nazionale di Ottica del Consiglio Nazionale delle Ricerche (CNR-INO), I-50019 Sesto Fiorentino, Italy}
\author{A. Trombettoni}
\affiliation{CNR-IOM DEMOCRITOS Simulation Center, I-34136 Trieste, Italy}
\affiliation{Scuola Internazionale Superiore di Studi Avanzati (SISSA), I-34136 Trieste, Italy}
\author{P. Cappellaro}
\affiliation{Department of Nuclear Science and Engineering, Massachusetts Institute of Technology, Cambridge, MA 02139}
\author{N. Fabbri}\email{fabbri@lens.unifi.it}
\affiliation{European Laboratory for Non-linear Spectroscopy (LENS), and Department of Physics and Astronomy, Universit\`a di Firenze, I-50019 Sesto Fiorentino, Italy}
\affiliation{Istituto Nazionale di Ottica del Consiglio Nazionale delle Ricerche (CNR-INO), I-50019 Sesto Fiorentino, Italy}

\begin{abstract}
Elucidating the energy transfer between a quantum system and a reservoir is a central issue in quantum non-equilibrium thermodynamics, which could provide novel tools to engineer quantum-enhanced heat engines. The lack of information on the reservoir inherently limits the practical insight that can be gained on the exchange process of open quantum systems. Here, we investigate the energy transfer for an open quantum system in the framework of quantum fluctuation relations.
As a novel toolbox, we employ a nitrogen-vacancy center spin qubit in diamond, subject to repeated quantum projective measurements and a tunable dissipation channel.
In the presence of energy fluctuations originated by dissipation and quantum projective measurements, the experimental results, supplemented by numerical simulations, show the validity of the energy exchange fluctuation relation, where the energy scale factor encodes missing reservoir information in the system out-of-equilibrium steady state properties. This result is complemented by a theoretical argument showing that, also for an open three-level quantum system, the existence of an out-of-equilibrium steady state dictates a unique time-independent value of the energy scale factor for which the fluctuation relation is verified.
Our findings pave the way to the investigation of energy exchange mechanisms in arbitrary open quantum systems.
\end{abstract}

\maketitle

\section{Introduction}
The connection between statistical properties of out-of-equilibrium dynamical systems, thermodynamics quantities and information theory has been deeply investigated in
classical and quantum systems and codified in terms of fluctuation relations~\cite{Evans93,Gallavotti95,Esposito09,Campisi11,Parrondo15,Goold16,Livi17}.
However, in open quantum systems, despite several contributions~\cite{Campisi09,Kafri12,Rastegin13,Albash13,Manzano15,Aberg18}, such connection is far from being completely understood, especially regarding the competition between thermal and quantum fluctuations. The latter assumes a paramount role at the nanoscale,
for example, for developing quantum thermal engines~\cite{Klatzow19,vonLindenfels19} or studying information--energy conversion~\cite{Toyabe10,Koski14}.
Accounting for the statistical fluctuations is the key to reformulate the second law of thermodynamics, usually expressed as inequalities, in terms of equalities. As a major example, the Jarzynski equality~\cite{Jarzynski97,Crooks99} relates the exponentiated negative work done on a system, averaged over a statistically relevant ensemble of realizations of the system dynamics, with the change in the free energy between two equilibrium thermal states. This framework has been also extended to describe the transport of energy and matter between different systems with different temperatures and chemical potentials~\cite{Jarzynski04, Andrieux09}. Remarkably, these relations hold for any kind of process driving the system arbitrarily far from equilibrium, provided that the initial state is in thermodynamic equilibrium.

In quantum mechanical settings, the fluctuation relations can be recast in terms of the characteristic function --- Fourier transform of the probability distribution function --- of the considered non-equilibrium quantity. This contains the full information about the fluctuations statistics and is obtained from two-time quantum correlations rather than by a single-time expectation value~\cite{Talkner07}. Still, in the absence of a heat reservoir, where the internal energy variation $\Delta E$ is solely due to work, a formally-equivalent quantum version of the Jarzynski equality (QJE)~\cite{Mukamel03,Tasaki00,Kurchan00} has been verified in various experimental settings with no heat flux involved, ranging from single trapped ions~\cite{An15} to liquid-state nuclear magnetic resonance platforms~\cite{Batalhao14}, atom chips~\cite{Cerisola17}, and superconducting Xmon qubits~\cite{Zhang18}.
\begin{figure}[t]
\begin{center}
\includegraphics[width=0.45\textwidth]{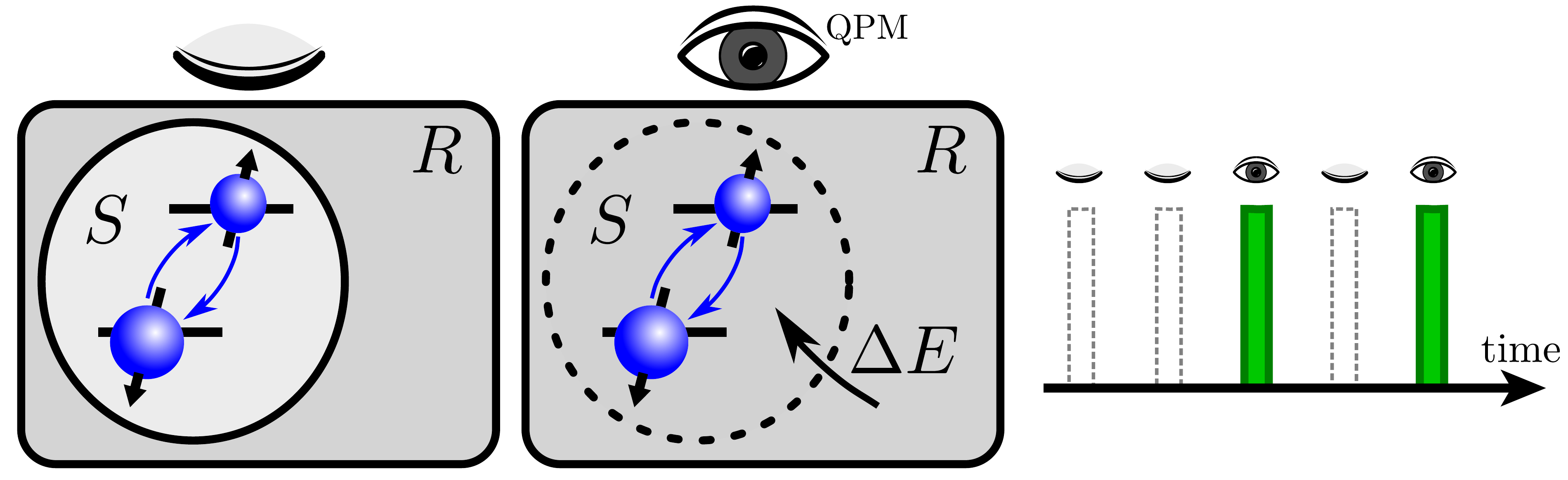}
\caption{{\bf Schematic illustration of the experiment.} A two-level system, in our case a single NV center qubit in diamond,  is subject to series of quantum projective measurements (QPMs), and controllable
energy exchange with a reservoir. We investigate the energy exchange fluctuations occurred in the system from the beginning to the end, over  a statistically relevant ensemble of realizations of the protocol.}
\label{fig:cartoon}
\end{center}
\end{figure}

Opening an energy exchange channel from a quantum system to a reservoir poses challenges for describing the nonequilibrium thermodynamic processes that take place~\cite{Campisi11}. While a dephasing dynamics can be effectively traced back to the case of an isolated system~\cite{Smith18}, the energy transport between a system and its environment is expected to be described by an exchange fluctuation relation~\cite{Esposito09,Campisi09,Campisi11}. Such relation was originally introduced to describe the heat exchange between two bodies in thermal contact~\cite{Jarzynski04}, each initially in thermal equilibrium. Evaluating the exchange fluctuation relation would then require information on the reservoir that is often not practically available.
Since the practically accessible quantity is the system energy variation at two different times~\cite{Batalhao14,Smith18,Zhang18}, the exchange fluctuation relation is conveniently expressed in terms of the characteristic function of the energy variation statistics $G(\varepsilon)\equiv \langle\exp{(-\varepsilon \Delta E)}\rangle$.
Thus, we argue that the energy exchange fluctuation relation for an open quantum system with a time-independent Hamiltonian can be written as
\begin{equation}
G(\varepsilon)  =1 \,,
\label{eq:energyExchange}
\end{equation}
where the scale factor $\varepsilon$ has to be determined and is expected to be of the order of the system energy.
Exemplarily, for a classical or quantum system under thermalizing dynamics, the energy scale $\varepsilon$ is predicted to be related to the inverse temperatures $\beta_{\rm in}$ and $\beta_{\infty}$ of the initial and final states as $\varepsilon=\Delta\beta$, with $\Delta\beta=\beta_{\rm in}-\beta_{\infty}$~\cite{Ramezani18}.
For a {\it quantum} system coupled to a tunable and generically non-thermal environment, it is not yet specified whether $\varepsilon$ such that $G(\varepsilon)=1$ exists, and whether and how it is related to the system dynamics and its asymptotic behavior. Here we investigate these issues.

We experimentally test the energy exchange fluctuation relation in a quantum simulator, as sketched in Fig.~\ref{fig:cartoon}, subject to repeated quantum projective measurements (QPMs) and energy dissipation, where the resulting dissipative dynamics drives the system towards an (out-of-equilibrium) energy steady state.
We realize the simulator with the use of a single nitrogen-vacancy~(NV) center qubit in diamond at room temperature, in the presence of trains of short laser pulses.
Each absorbed laser pulse results in a QPM~\cite{Wolters13}, and in an energy redistribution that can be modeled as a controlled energy exchange with a Markovian reservoir~\cite{Klatzow19}. The time intervals between QPMs follow a stochastic distribution due to the finite absorption probability. Tuning the laser duration and power enables the control of the coupling strength between the quantum system and the reservoir.

The combined effect of QPMs and dissipation can create or destroy quantum coherence during the system dynamics, an effect that goes beyond the classical description.
While QPMs affect the energy distribution of the system, they are expected to preserve the validity of quantum fluctuation relations~\cite{Campisi10,Campisi11b}, also for stochastic distributions of QPMs~\cite{Gherardini18}. However, the energy fluctuations of a quantum system in the presence of QPMs and dissipative dynamics have not been studied yet. Measuring the statistics of the exponentiated energy fluctuations through a two-point measurement (TPM) protocol, we experimentally verify the exchange fluctuation relation for an open two-level quantum system. We find out a uniquely determined value of $\varepsilon$ for which $G(\varepsilon)=1$,  incorporating missing information about the reservoir. For a two-level system, where any diagonal density matrix in the energy basis can be recast in terms of an effective temperature, $\varepsilon$ encompasses the initial and final populations of the quantum system through their effective temperatures.

While these results have been obtained for an effective quantum two-level system, we provide a further analysis involving a three-level system that is affected by a spontaneous emission process. In the Appendix 2 we show with this example that the validity of the exchange fluctuation relation is conditioned by the existence of a unique non-trivial time-independent energy scale factor $\varepsilon$, without requiring thermalizing dynamics. Therefore, our results are representative of quantum systems with dimension larger than two subjected to dissipation dynamics.

\begin{figure*}[]
\begin{center}
\includegraphics[width=0.9\textwidth]{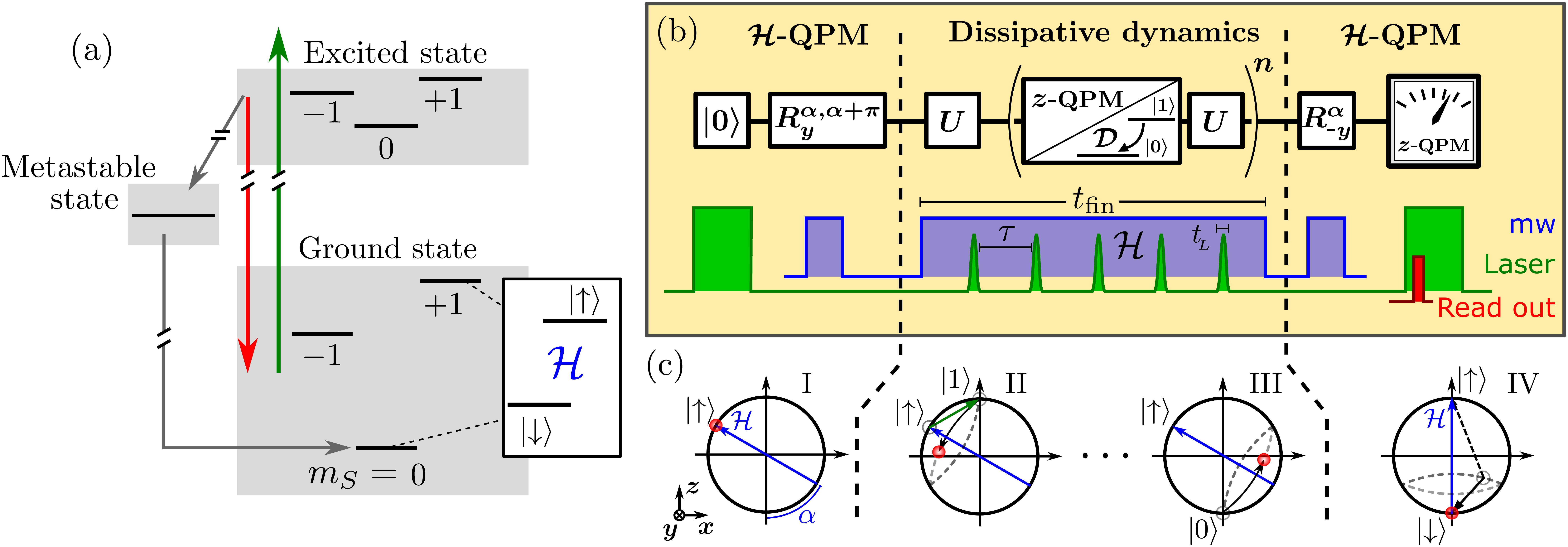}
\caption{
{\bf Protocol implementation.} (a) Schematic representation of the photodynamics of a nitrogen-vacancy  center.
Photon absorption (green arrow) and  spin preserving spontaneous emission (red arrow) between the electronic ground and excited levels realize quantum projective measurements of the spin state along the $z$ axis ($z$--QPM). A non-radiative decay channel (gray arrows) enables controlled optical spin pumping to the $\ket{0}$ state. Inset: Effective two-level system considered in the protocol, formed by two dressed spin states set by quasi-resonant continuous driving (see Eq.~\eqref{eq:H}). (b) Block-diagram of the two-point measurement protocol implementation, and experimental pulse sequence. After initialization in the $\ket{0}$ state, for each protocol repetition a Hamiltonian eigenstate $\ket{\uparrow}$ ($\ket{\downarrow}$) is prepared by applying a rotating microwave (mw) gate $R_y^\alpha$ ($R_y^{\alpha+\pi}$). During the  time $t_{\rm fin}$, the mw-driven spin unitary evolution set by the Hamiltonian $\mathcal{H}$, defined in Eq.~\eqref{eq:H}, is perturbed by equidistant short laser pulses  acting as $z$--QPMs plus a dissipation channel ($\mathcal{D}$). The inter-pulse time ($270 \le \tau \le 750$~ns) is much longer than each pulse duration ($t_L=41$~ns). At the end, a quantum projective measurement of the final state energy ($\mathcal{H}$--QPM) is realized with a mw gate  $R_{-y}^\alpha$ and a spin selective fluorescence intensity measurement -- read out.
(c) Exemplary spin state evolution (red dots) on the Bloch sphere for a single realization of the protocol.
(c.I) Initially prepared Hamiltonian eigenstate, e.g.\,$\ket{\uparrow}$. (c.II) $\ket{\uparrow}$  is projected in one of the $\sigma_z$ eigenstates, e.g.\,$\ket{+1}$, due to photon absorption as denoted by the the green arrow, and then evolves under unitary dynamics (dashed circle)  until a subsequent photon absorption.
(c.III) Due to the spin amplitude damping, after several laser absorptions the system would most likely be in the $\ket{0}$ state. After the last short laser absorption the state follows a unitary evolution. In average over several realizations of the protocol, the state before the $\mathcal{H}$--QPM has the same energy than the state $\ket{0}$, but with an unknown phase in the Hamiltonian basis. (c.IV) Applying a rotation $R_{-y}^\alpha$ allows us to obtain the energy of the system by measuring $\sigma_z$. Indeed, the final state is projected into one of the Hamiltonian eigenstates (e.g. $\ket{\downarrow}$).
}
\label{fig:LevelsAndProtocol}
\end{center}
\end{figure*}

\section{Protocol implementation}	
\begin{figure}[t]
\begin{center}
\includegraphics[width=0.975\columnwidth]{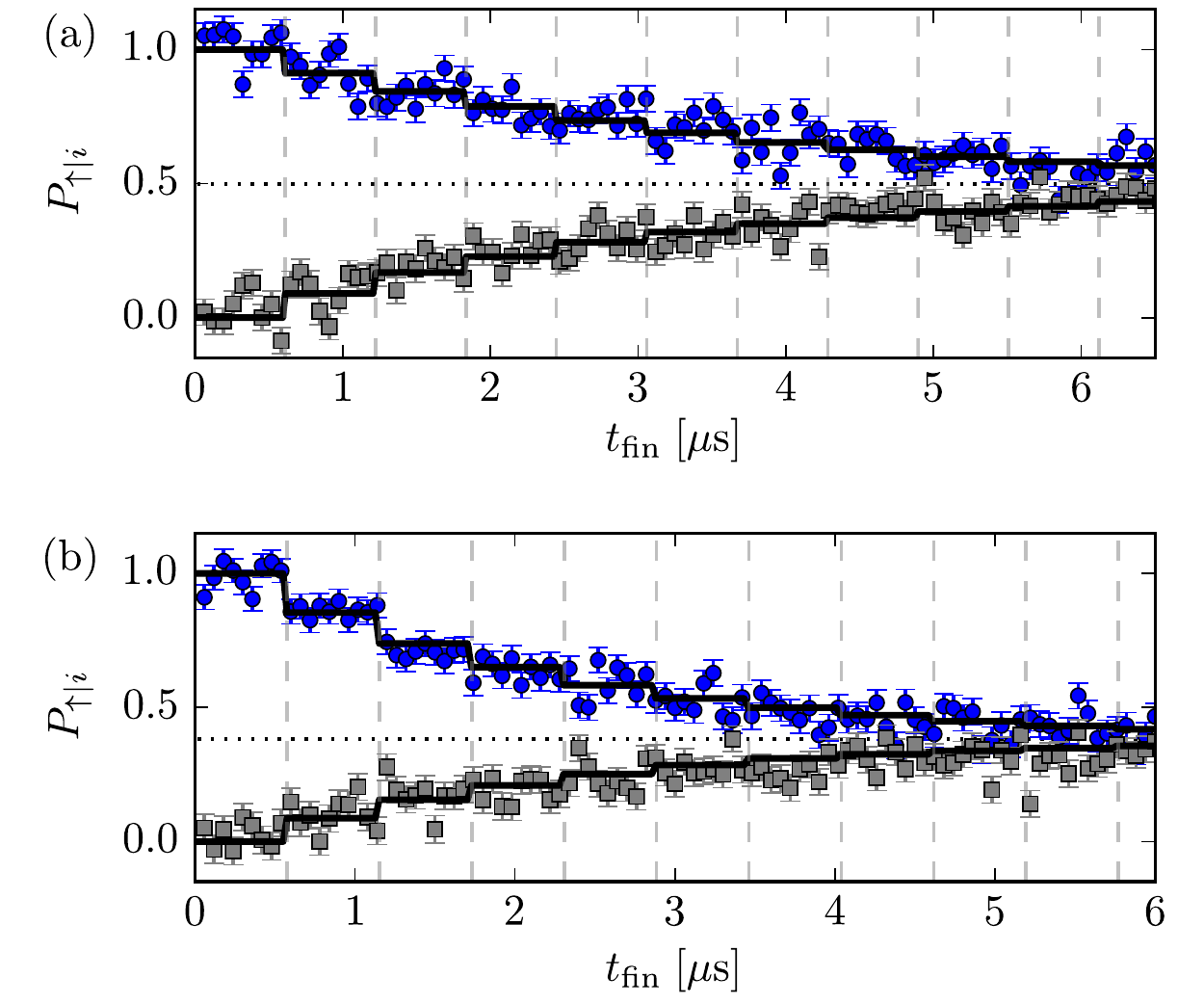}
\caption{
{\bf Energy jumps of the qubit  under projective measurements and dissipative dynamics.} Conditional probabilities $P_{\uparrow|\uparrow}$ and $P_{\uparrow|\downarrow}$ to measure $E_{f}=E_{\uparrow}$ as a function of the evolution time $t_{\rm fin}$ starting from the Hamiltonian eigenstates $\ket{\uparrow}$ (blue dots) or $\ket{\downarrow}$ (grey dots), with $\alpha = \pi/2$ (a) and $\alpha = \pi/3$~(b). Vertical grids indicate the time position of each laser pulse, separated by $\tau=5\pi/3\omega$. The probability that a $z$--QPM occurs during each laser pulse is $0.18$. Error bars denote the experimental uncertainty given by the photon shot noise. The solid lines are obtained by a Monte Carlo simulation of the dynamics (see Sec.~S.II from Supplemental Material).
The horizontal dotted lines represent the conditional probabilities for the asymptotic (out-of-equilibrium) steady states (see text).
}
\label{fig:JarzyEnergy}
\end{center}
\end{figure}

Here we introduce our experimental platform and describe the protocol used for measuring the energy  fluctuations. We show that this protocol can be broadly applied to any finite-dimensional quantum system, and we detail it for our specific experimental setup.

The experimental platform is based on a negatively charged NV center --- a localized impurity in the diamond lattice based on a nitrogen substitutional atom and a nearby vacancy --- which forms an electronic spin $S=1$ in its orbital ground-state (see Fig.~\ref{fig:LevelsAndProtocol}(a)). The electronic spin can be  initialized into the state $\ket{0}$, where $\ket{m_s=0,\pm1}$ stands for the eigenvalues of the spin operator $S_z$ along the NV symmetry axis, via optical spin pumping under laser excitation~\cite{Doherty13}.
A strong magnetic bias field removes the degeneracy of the spin states $\ket{\pm1}$, allowing selective coherent manipulation of the transition $\ket{0}\leftrightarrow\ket{+1}$.
The spin-dependent intensity of the emitted fluorescence enables the optical readout of the states $\ket{0}$ and $\ket{+1}$.
A continuous nearly-resonant microwave field sets the Hamiltonian $\mathcal{H}$ of the two-level system in the frame rotating at the microwave frequency,
\begin{equation}
\mathcal{H}=\frac{\hbar \omega}{2} (\cos\!\alpha \,\,\sigma_z - \sin\!\alpha\,\,\sigma_x ),
\label{eq:H}
\end{equation}
with eigenstates $\{\ket{\uparrow},\ket{\downarrow}\} = \{\cos\frac{\alpha}{2}\ket{0} - \sin\frac{\alpha}{2}\ket{+1}$, $\sin\frac{\alpha}{2}\ket{0} + \cos\frac{\alpha}{2}\ket{+1}\}$, and eigenvalues $E_\uparrow=\hbar\omega/2$ and $E_\downarrow=-\hbar\omega/2$. Here, $\sigma_i$ are Pauli matrices, $\tan \alpha = -\Omega / \delta$ and $\omega = \sqrt{\delta^2 + \Omega^2}$, $\Omega=1.3$~MHz being the bare Rabi frequency, and  $\delta\in[0,\Omega]$ the microwave detuning.
The Hamiltonian~\eqref{eq:H} remains unchanged during the protocol, and the non-unitary dynamics is due to repeated QPMs and dissipation acting along the $\sigma_z$ axis, as explained below.

The resulting quantum dynamics induces energy fluctuations of the spin system. To characterize the statistics of these energy fluctuations, we employ a TPM protocol, where the energy is measured at the initial and final times of the process. We implement this protocol as shown in Fig.~\ref{fig:LevelsAndProtocol}\,(b) and specified below.
Figure~\ref{fig:LevelsAndProtocol}(c) shows an example of a single trajectory followed by the system during the complete protocol.

{\it(i) Preparation of initial thermal states, and energy quantum projective measurement ($\mathcal{H}$--QPM)}.
Each realization starts by preparing the system in one of its energy eigenstates $\ket{i}\!\!\bra{i}$. The experiment is then repeated $P_i N$ times for each eigenstate over a statistical ensemble of $N$ realizations, where $P_i$ is the population fraction of each energy eigenstate following a thermal distribution~\footnote{Note that for system dimension higher than two, not all the probabilistic mixtures of energy eigenstates correspond to a thermal state.}.
Independently of the system dimension, this is equivalent to considering the statistical result of an initial energy measurement ($\mathcal{H}$--QPM) applied to the thermal state $\sum_i^n P_i \ket{i}\!\!\bra{i}$, provided that $N$ is large enough.

In the experiment, each of the two energy eigenstates (described by the density operators $\varrho_\uparrow=\ket{\uparrow}\!\!\bra{\uparrow}$ and $\varrho_\downarrow=\ket{\downarrow}\!\!\bra{\downarrow}$) is prepared by optically pumping the NV into the $\ket{0}$ state and then applying a rotation $R^\alpha$~({\it resp.,} $R^{\alpha+\pi}$) along $\sigma_y$, via a microwave (mw) gate as depicted in Fig.~\ref{fig:LevelsAndProtocol}(b).

{\it(ii) Evolution under repeated quantum projective measurements and dissipative dynamics.} The system is repeatedly opened for short time intervals and stays closed otherwise, due to being subject to a series of QPMs of an operator non-commuting with the Hamiltonian, and a dissipative dynamics. These effects on top of the unitary evolution $U=e^{-i\mathcal{H}t/\hbar}$ lead the system into an asymptotic steady state.

In the experiment, we apply to the NV center trains of short laser pulses with duration $t_L$ at intervals $\tau$, as depicted in Fig.~\ref{fig:LevelsAndProtocol}(b). The laser pulses trigger cycles of spin-preserving radiative transitions from the ground to the excited states (see Fig.~\ref{fig:LevelsAndProtocol}(a) and Appendix 1). Upon photon absorption, any superposition or mixed spin state is projected into either one of the two $\sigma_z$ eigenstates $\ket{0}$ or $\ket{+1}$, while the state coherence imprinted by the microwave during the prior evolution is destroyed in the $\sigma_z$ basis. This results in a quantum projective measurement of $\sigma_z$  ($z$--QPM)~\cite{Wolters13}, even when the measurement outcome is not recorded. Significantly, this mechanism produces coherence in other bases, such as the energy basis (for $\alpha\neq0$), as shown in Fig.~\ref{fig:LevelsAndProtocol}(c).
Due to the finite photon-absorption probability, a train of equidistant laser pulses entails a stochastic time distribution of $z$--QPMs, for each single realization.
In addition, the absorption of laser pulses induces a partial population transfer to $\ket{0}$, due to optical pumping (see Fig.~\ref{fig:LevelsAndProtocol}(a)), with an effective rate that depends on the number of excitation-decay cycles performed by the system, which can be controlled by changing the laser pulse duration and power.
This spin amplitude damping mechanism is equivalent to a controlled dissipative channel towards $\ket{0}$ in the two-level system, which together with $z$--QPMs, incorporates all the laser-induced NV photo-dynamics involving the ground and excited triplet states, and the metastable singlet state.
The overall effect takes the system into an asymptotic out-of-equilibrium steady-state in the Hamiltonian basis.

{\it(iii) Measurement of the final energy.}
The statistics of $\Delta E$ is provided by the conditional probabilities $P_{j|i} = P(E_{\rm fin} = E_j | E_{\rm in} =E_i)$
to measure $E_j$ as the final energy, once known the initial energy $E_i$ (see Eq.~\eqref{eq:probsDeltaE} of the Appendix).
For an $n$-dimensional system,  measuring $n - 1$ diagonal elements of the final density operator
for each initial energy eigenstate $|i\rangle\!\langle i|$
bestows a full knowledge of all the conditional probabilities.
For the NV qubit, we measure the conditional probabilities $P_{\uparrow|\uparrow}$ and $P_{\uparrow|\downarrow}$
for the spin to go in $\varrho_\uparrow$ (with energy
$E_{\rm {fin}}=E_\uparrow$), when starting respectively from $\varrho_\uparrow$ or $\varrho_\downarrow$. To implement the final energy measurement ($\mathcal{H}$--QPM), we apply a mw gate that maps $\mathcal{H}$ into $\sigma_z$ (as detailed in Fig.~\ref{fig:LevelsAndProtocol}(b) and (c).IV), and then measure the $\sigma_z$ operator by detecting the presence (or absence) of emitted photons. Low collection efficiency and photon shot noise impose the need of repeating the procedure several times ($\sim 1.6 \times 10^6$) and averaging over the detected intensity to reduce the readout uncertainty.

\section{Statistics of the energy variation}
Figure~\ref{fig:JarzyEnergy} shows the conditional probabilities $P_{\uparrow|\downarrow}$ and $P_{\uparrow|\uparrow}$ as a function of the evolution time $t_{\rm fin}$,  obtained in the  experiment. The competing effects of $z$--QPMs and dissipation lead to a non-trivial dynamics, affecting the energy fluctuation distribution.
To quantitatively support that the considered two-level model provides an accurate description of the system dynamics, we performed a numerical Monte Carlo simulation of the dynamics (see Sec.~S.II from Supplemental Material) and we found excellent agreement with data. Note that the only fit parameter is the absorption probability, which depends on the laser power and characterizes the stochasticity of the protocol. In the absence of laser pulses, the spin qubit is a closed system and the energy eigenstates do not evolve in time (usually referred to as \emph{spin lock}), while the absorption of laser pulses produce discrete energy jumps.

\begin{figure}
\begin{center}
\includegraphics[width=0.925\columnwidth]{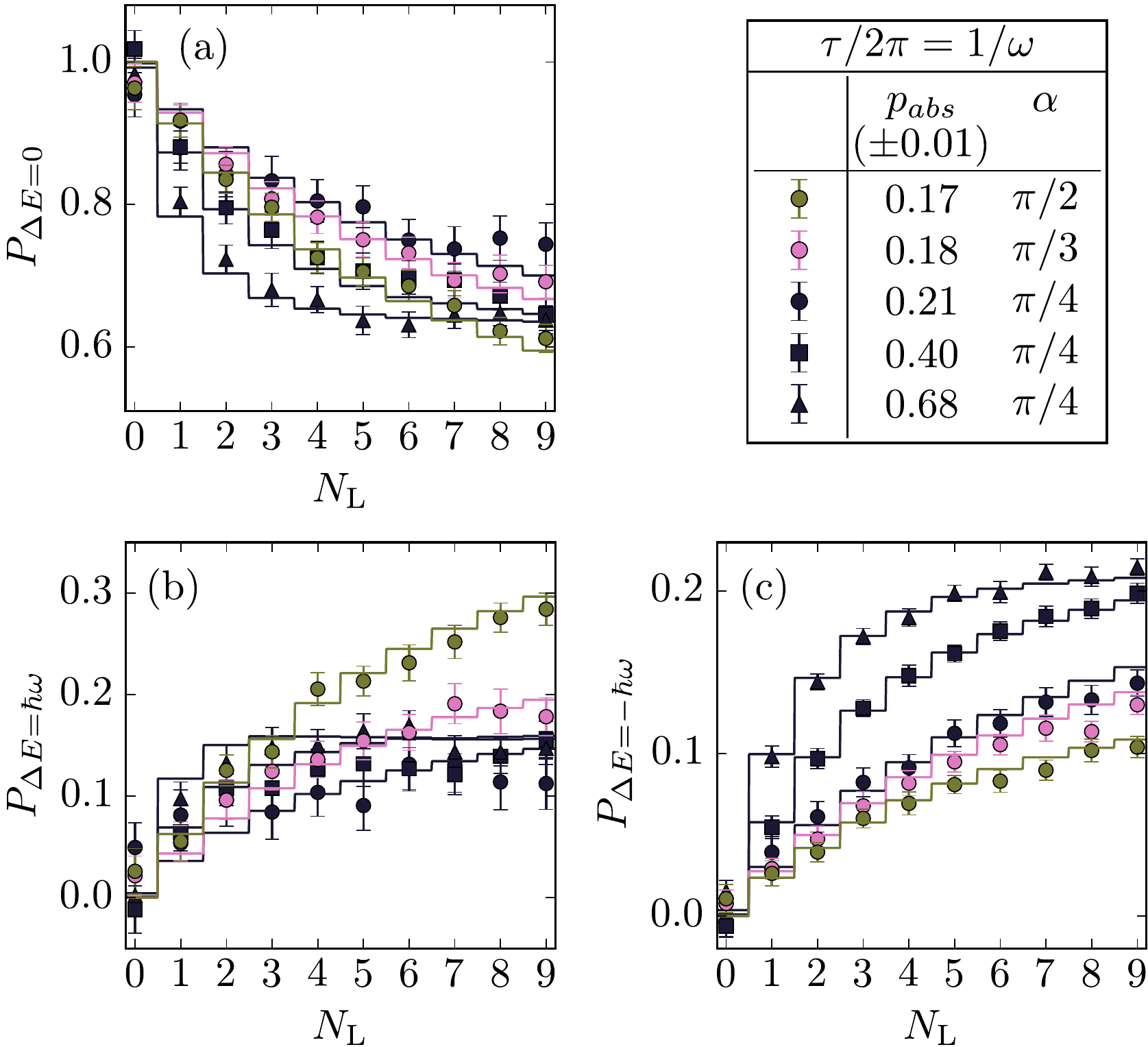}
\end{center}
\caption{
{\bf Statistics of the energy variation under different protocols.} Probability to obtain $\Delta E = 0,\,+\hbar\omega,\,-\hbar\omega$, respectively, as a function of the number of laser pulses $N_L$ experienced by the qubit, with initial probability $P_\uparrow = 1/(1+e)$. For each considered $N_L$ value, the experimental points represent the average over $\sim 10$ different values of final evolution times $t_{\rm fin}$.
The error bars are due to the uncertainty on the measured photoluminescence intensity. The solid lines are the joint probabilities obtained from the numerical simulation of the system dynamics.
}
\label{fig:summary_QJE}
\end{figure}

The energy variation occurred to the qubit after the process can assume one of the three values $\Delta E \in \{-\hbar\omega,0,+\hbar\omega\}$.
Figure~\ref{fig:summary_QJE} shows the distribution of energy variation ($P_{\Delta E=0}$, $P_{\Delta E=+\hbar\omega}$, and $P_{\Delta E=-\hbar\omega}$), for a fixed initial mixed state when varying the value of $\alpha$ and the power of the laser pulses. This result shows that in the presence of $z$--QPMs and dissipation, the energy distribution of the quantum system is modified. The system jumps between states with different coherences in the energy basis -- as sketched in Fig.~\ref{fig:LevelsAndProtocol}(c) -- and finally reaches, for a large number of $z$--QPMs, an out-of-equilibrium steady state, which does not depend on the initial state.
The photon absorption probability dictates how fast the system approaches the asymptotic steady state.
Then, the final projective energy measurement returns, on average, a mixed state defined by the balance between the energy variation due to $z$--QPMs applied to the system and the dissipation channel.

For ideal equally-spaced $z$--QPMs (perfect absorption, and only spin-preserving optical transitions) the asymptotic probability to find the spin in the $\ket{\uparrow}$ state $P_\uparrow^\infty=P_{\uparrow|\uparrow}^\infty=P_{\uparrow|\downarrow}^\infty$ can be analytically computed by modeling the spin temporal evolution with a master equation in the Lindblad formalism, yielding
\begin{equation}
P_\uparrow^\infty= \frac{1}{2} \left(1 - \frac{\left(1-e^{-t_L\Gamma_{\mathcal{D}}}\right) \cos\!\alpha}{1-e^{-t_L\Gamma_{\mathcal{D}}} \mu(\alpha,\tau)}\right),
\end{equation}
with $\mu \equiv 1-2\,(\sin\!\alpha \,\sin\!\frac{\omega \tau}{2})^2$ (see Sec.~S.II from Supplemental Material). Given the experimental dissipation rate $\Gamma_{\mathcal{D}}$, the analytic prediction of $P_\uparrow^\infty$
matches the numerical simulations for ideal equally-spaced $z$--QPMs. In the experiment, the stochasticity of the temporal distribution of $z$--QPMs --- induced by the finite photon absorption --- removes the strong dependence on $\tau$ (see Sec.~S.III in the Supplemental Material). The analytical model is still a good approximation of the system dynamics, provided one replaces $\tau$ with an effective inter-pulse spacing (see Fig.~IV.S).

In the absence of dissipation ($\Gamma_{\mathcal{D}}=0$), $z$--QPMs bring the system into an equilibrium thermal state with infinite temperature~\cite{Yi11,Elouard17} ($P_\uparrow^\infty=1/2$). The cases $\alpha=\{0,\pi\}$ and  $\tau=2\pi/\omega$, {\it i.e.}, $\mu=1$, are exceptions~\footnote{If $\alpha=\{0,\pi\}$, $z$--QPMs do not affect the spin dynamics after the first energy measurement. For $\tau\omega=2\pi$, $U(\tau)$ is the identity operator, therefore only the first $z$--QPM affect the spin energy. Still, dissipation brings the spin state to the state $\ket{0}$.}.
The same asymptotic probability $P_\uparrow^\infty=1/2$ is observed also in presence of dissipation, when $z$--QPMs and amplitude damping act in a direction orthogonal to the Hamiltonian ($\alpha=\pi/2$, $\Gamma_{\mathcal{D}}>0$), as experimentally confirmed (see Fig.~\ref{fig:JarzyEnergy}(a)).
Indeed for $\alpha=\pi/2$ the asymptotic state before the final energy measurement is a fully coherent state in the Hamiltonian basis such that $\braket{\mathcal{H}}=0$, thus the density matrix after the final energy measurement corresponds, in average, to a completely mixed state (that is, a thermal state with infinite effective temperature). In this configuration ($\alpha=\pi/2$), during a TPM protocol, dissipative dynamics is indistinguishable from unitary dynamics plus repeated measurements ($\Gamma_{\mathcal{D}}=0$)~\cite{Campisi11b}.

\begin{figure*}
\begin{center}
\includegraphics[width=0.95\textwidth]{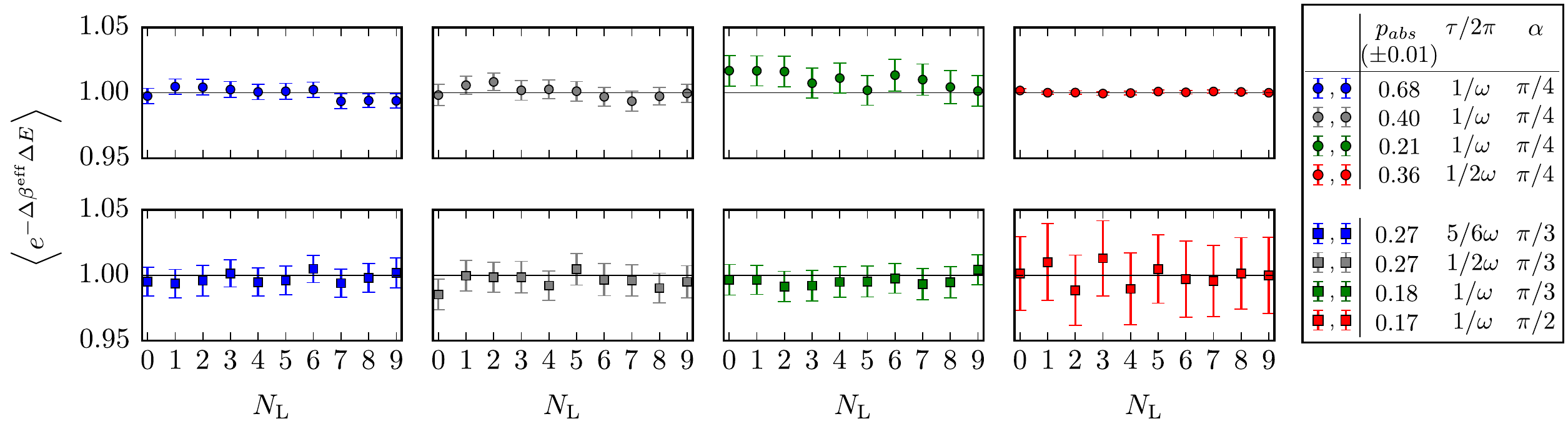}
\end{center}
\caption{
{\bf Verification of the exchange fluctuation relation for an open quantum system.} Experimental values of $\langle e^{-\Delta\beta^{({\rm eff})}\Delta E}\rangle$ (obtained as in Eq.~\eqref{eq:LHS} in Appendix 2) as a function of the number of laser pulses $N_L$. Each dataset represents a different combination between the parameters, the angle $\alpha$ (see Eq.~\eqref{eq:H}), the time $\tau$ between laser pulses, and the photon-absorption probability $p_{abs}$. These data constitute the experimental verification of Eq.~\eqref{eq:jarzy2}. The initial probability is set to $P_\uparrow = 1/(1+e)$, while the asymptotic probability $P_\uparrow^{\infty}$ is acquired from experimental data, as described in the Appendix 2. Both quantities define the value of $\Delta\beta^{({\rm eff})}$.
}
\label{fig5}
\end{figure*}

\section{Exchange fluctuation relation}
For two macroscopic systems $S_1$ and $S_2$ placed in thermal contact for a finite lapse of time, the statistics of exchanged heat $Q$ is known to be described by the heat exchange fluctuation relation $\langle e^{-(\beta_{{\rm in},S_1}-\beta_{{\rm in},S_2})Q}\rangle=1$, where $\beta_{{\rm in},S_1}$ and $\beta_{{\rm in},S_2}$ are the initial inverse temperatures of $S_1$ and $S_2$ and $Q=\Delta E_{S_1}=-\Delta E_{S_2}$~\cite{Jarzynski04}. For thermalizing quantum dynamics, the non trivial value of $\varepsilon$ for which Eq.~(\ref{eq:energyExchange}) is verified is $\varepsilon=\Delta\beta \equiv \beta_{\rm in} - \beta_\infty$, where $\beta_{\rm in}$ and $\beta_\infty$ are the initial and asymptotic inverse temperatures of the system~\cite{Ramezani18}. These two results are equivalent in the case of a quantum system interacting with a thermal reservoir at inverse temperature $\beta_\infty$. However, the interaction with a non-thermal reservoir calls for a deeper understanding of the role played by the energy scale factor $\varepsilon$. With this aim, we implemented a numerical simulation to show (see Appendix) that, for a three-level quantum system asymptotically approaching a steady state in the energy basis (SSE), there exists one single time-independent $\varepsilon \neq 0$ such that $\langle\exp{(-\varepsilon \Delta E)}\rangle = 1$.
The system is brought into this steady state with constant energy in time, which does not depend on the initial state, by means of a dissipation channel that is modeled as a decay induced by spontaneous emission. After the second energy measurement of the TPM protocol, the quantum system is in a mixed state that is not necessarily thermal. This means that the effects of the dissipative channel on a quantum system may not be modeled by the effective interaction of the system with a thermal reservoir. However, also in this case, the validity of Eq.~(\ref{eq:energyExchange}) is ensured and, specifically, related to the existence of a unique non-trivial constant value of $\varepsilon$ when the system is at the SSE. It is worth noting that $\varepsilon$ depends on the populations of the density matrix of both the initial state and the steady state. In the special case that the mixed state after the TPM protocol is thermal at inverse temperature $\beta_\infty$, the fluctuation relation $\langle\exp(-\Delta\beta\Delta E)\rangle = 1$ is recovered~\cite{Ramezani18}. The above considerations show that quantum fluctuation relations hold also in the steady state regime~\cite{Andrieux09}, while in the transient regime a model-dependent behaviour is expected. We observed that the energy scale $\varepsilon$ does not depend on the non-unitary map defined by the applied protocol, but only depends on the asymptotic steady state.

In the experiment, the initial state is described by a density matrix diagonal in the energy basis and, since the system is effectively two-dimensional, can be written as a thermal state with an effective inverse temperature $\beta_{\rm in}^{({\rm eff})}$. Similarly, the stationary mixed state after the second energy measurement of the TPM protocol is, in average, equivalent to a thermal state with effective inverse temperature $\beta_{\infty}^{({\rm eff})}$ (see Eq.~\eqref{eq:thermalrho} of Appendix).
Notice that for each single realization and before the second energy measurement, the state evolves in time even in the asymptotic limit, marking a difference between dissipative and thermalizing dynamics for a two level system. Figure~\ref{fig5} shows that the experimental data and simulation always verify the relation
\begin{equation}
\langle\exp{(- \varepsilon \Delta E)}\rangle =1\,,
\label{eq:jarzy2}
\end{equation}
with
\begin{equation}
\varepsilon= \beta_{\rm in}^{({\rm eff})}-\beta_{\infty}^{({\rm eff})} \equiv \Delta\beta^{({\rm eff})},
\label{eq:jarzy2bis}
\end{equation}
irrespective of the initial state and the applied protocol, i.e., relative orientation between the $z$--QPM operator and the system Hamiltonian, inter-pulse time intervals and photo-absorption probability.
We emphasize that, albeit our measurements are done for a quantum two-level systems and therefore can be interpreted in terms of effective temperature, the formalism and the conclusion are expected to hold for a generic quantum system, including multi-level systems, as predicted by the numerical example
in the Appendix.

\section{Conclusions}
We explored the quantum exchange fluctuation relation for an open quantum system coupled to a tunable dissipative channel. We investigated the interplay between quantum projective measurements and a dissipation channel, and we proposed a formulation of the energy exchange fluctuation relation,
where the energy scaling factor depends on the populations of the stationary density matrix  reached by the open quantum system. We showed that this formulation holds also for a three-level system that asymptotically reaches a steady state in the energy basis, suggesting that this result might be extended to a general finite-dimensional system.
For the implemented two-level system, the energy scaling factor can be formulated in terms of the effective temperatures of the initial and final states of the system. At the steady state, the final effective temperature is indeed an invariant quantity, irrespective of the initial state. We have shown that this exchange fluctuation relation holds for any direction, with respect to the Hamiltonian, along which the intermediate quantum projective measurements are applied. In addition, we have observed that the exchange fluctuation relation is robust against the presence of randomness in the time intervals between measurements, as theoretically predicted in \cite{Gherardini18}.
Our experimental study is enabled by the use of a single NV center in diamond at room temperature.  We exploit the high control on the spin degrees of freedom, under the effect of trains of short laser pulses that perform quantum projective measurements and controllably open the two-level system, through a dissipation channel whose interaction coupling with the external surroundings can be tuned. 
This work, therefore, exploits NV centers in diamond as a quantum simulator to explore the physics of an out-of-equilibrium open quantum system, and to verify quantum fluctuation relations.

Our work paves the way for the investigation of Jarzynski-like equalities for general open quantum systems beyond two-dimensional Hilbert spaces.
In addition, our results consolidate NV centers as a suitable platform to study phenomena related to open quantum systems, for example, to study the role of coherence in energy transport, or to experimentally verify different quantum fluctuation relations~(QFR), such as the QFR for engines or refrigerators~\cite{Campisi15}, or the so called generalized QFR~\cite{Sagawa10}. 
Finally
, we hope that our work will stimulate further research to experimentally test our findings with other physical realizations, ranging from ions~\cite{An15} to superconducting devices~\cite{Koski13,Zhang18}, and to ultracold gases~\cite{DeChiara15}. Furthermore, our results can contribute to the implementation of heat engines working in out-of-equilibrium regimes~\cite{Klatzow19}.


\section*{Acknowledgements}
This work was supported by EU-FP7 ERC Starting Q-SEnS2 (Grant No. 337135), Fondazione CR Firenze through the project Q-BIOSCAN, and by the MISTI Global Seed Funds MIT-FVG Collaboration Grant ``NV centers for the test of the Quantum Jarzynski Equality (NVQJE)''. We thank A. Sone, S. Ruffo, and M. Campisi for fruitful discussions, and M. Inguscio for continuous inspiration and support.

\section*{Appendix: Methods}

\subsection*{Experimental platform and modeling}
We used a single NV center hosted in an electronic grade diamond sample, with 14-N concentration $< 5$~ppb (Element Six). The NV center is optically addressed at ambient conditions with a home-built confocal microscope and its electronic spin is manipulated via resonant microwave driving. The NV center is chosen to be free from proximal 13-C nuclear spins. The 14-N spin is polarized due to a static bias magnetic field of 394~G, combined with electronic spin pumping~\cite{Poggiali17}. The long coherence time of the nuclear spin ensures that it remains unaffected during the experiment. A microwave coherently manipulate the effective two-level system, composed by the $m_s=0$ and $m_s=+1$ levels of the ground state. On the experiment timescales ($\sim\mu$s) spin-lattice relaxation is negligible ($T_1 \sim $~ms), while the Rabi driving prevents spin dephasing to occur~\cite{MacQuarrie15}.

The absorption of $532$~nm laser light pulses excites the NV-center electronic spin from the ground to the excited triplet states.
The decay involves (i) radiative transitions to the ground state, spin-preserving ($\sim 96.5$~\%, see Sec.~S.I from Supplemetal Material), generating a red-shifted photoluminescence with zero-phonon line at $637$~nm,
and (ii) non-radiative transitions through a singlet metastable state.
Thus, the interaction with short laser pulses has a probability $(1-p_\mathrm{diss})$ to result in an ideal $z$--QPM, but also a finite probability ($p_\mathrm{diss}<1$) to destroy the state and force the resulting state to be $\ket{0}$, this process gives origin to the dissipative dynamics. Even when we cannot completely isolate each of these two effects, by changing the short laser pulses duration and intensity, we can control the value of $p_\mathrm{diss}$ to be between~$\simeq 0.44$ and $1$.
The photodynamics of the NV center is thus well described with a seven-level model~\cite{Manson06, Wolters13}.
However, our experiments can be well-reproduced by an effective two-level model. The neglected photodynamics occurring through the hidden physical states is re-absorbed through an effective photon absorption probability $p_{abs}$, an effective dissipation probability $p_{abs}\Gamma_{\mathcal{D}}t_L$ and a correction on the $z$--QPM  outcome that takes into account the non-spin conserving probability.  In this regard, the simulations shown in the main text were realized by using a two-level system with absorption probability in the range $18$ - $68$~\%, and $44$~\% conditional probability to move populations to $\ket{0}$ once projected in $\ket{+1}$.
The results of the analysis with a seven-level model and its comparison with the two-level one are reported in Sec.~S.II of the Supplemental Material.

\subsection*{Energy steady state regime enabling fluctuation relations: a three-level system case-study}

From the analysis of the experimental data we have observed the connection between the stationary state SSE of a two-level quantum system and the existence of a unique, finite, time-independent value of $\varepsilon$, obeying the fluctuation relation $G(\varepsilon)=\langle\exp\left(-\varepsilon\Delta E\right)\rangle=1$.
In this section, we extend this result to to an exemplary case of a three-level system (3LS) subjected to dissipative dynamics.

Specifically, we have chosen a 3LS governed by the following Hamiltonian:
\begin{align}
H &= \omega_{12}\left( |1\rangle\!\langle 2| + |2\rangle\!\langle 1|\right) + \omega_{13}\left( |1\rangle\!\langle 3| + |3\rangle\!\langle 1|\right)  \notag \\
&= \begin{pmatrix} 0 & \omega_{12} & \omega_{13} \\ \omega_{12} & 0 & 0 \\ \omega_{13} & 0 & 0 \end{pmatrix},
\end{align}
where $|j\rangle$ denotes the $j-$th level of the system, and $\omega_{12}$, $\omega_{13}$ are the coupling rates between the first level and the second and the third ones. The system is also characterized by a decay channel (mimicking a spontaneous emission process) with rate $\Gamma$ between the first and second level of the 3LS. We describe the system dynamics by means of a Lindblad master equation, whereby the spontaneous decay is given by the following Lindbladian super-operator:
\begin{equation}
L = \sqrt{\Gamma}|1\rangle\!\langle 2| =
\begin{pmatrix}
0 & \sqrt{\Gamma} & 0 \\ 0 & 0 & 0 \\ 0 & 0 & 0
\end{pmatrix}.
\end{equation}
Note that,
albeit no intermediate QPMs are applied to the quantum system, the dynamics of the 3LS is dissipative due to the presence of a spontaneous decay term, here modelled by means of the Lindbladian formalism.

\begin{figure}
\begin{center}
\includegraphics[width=0.49\textwidth]{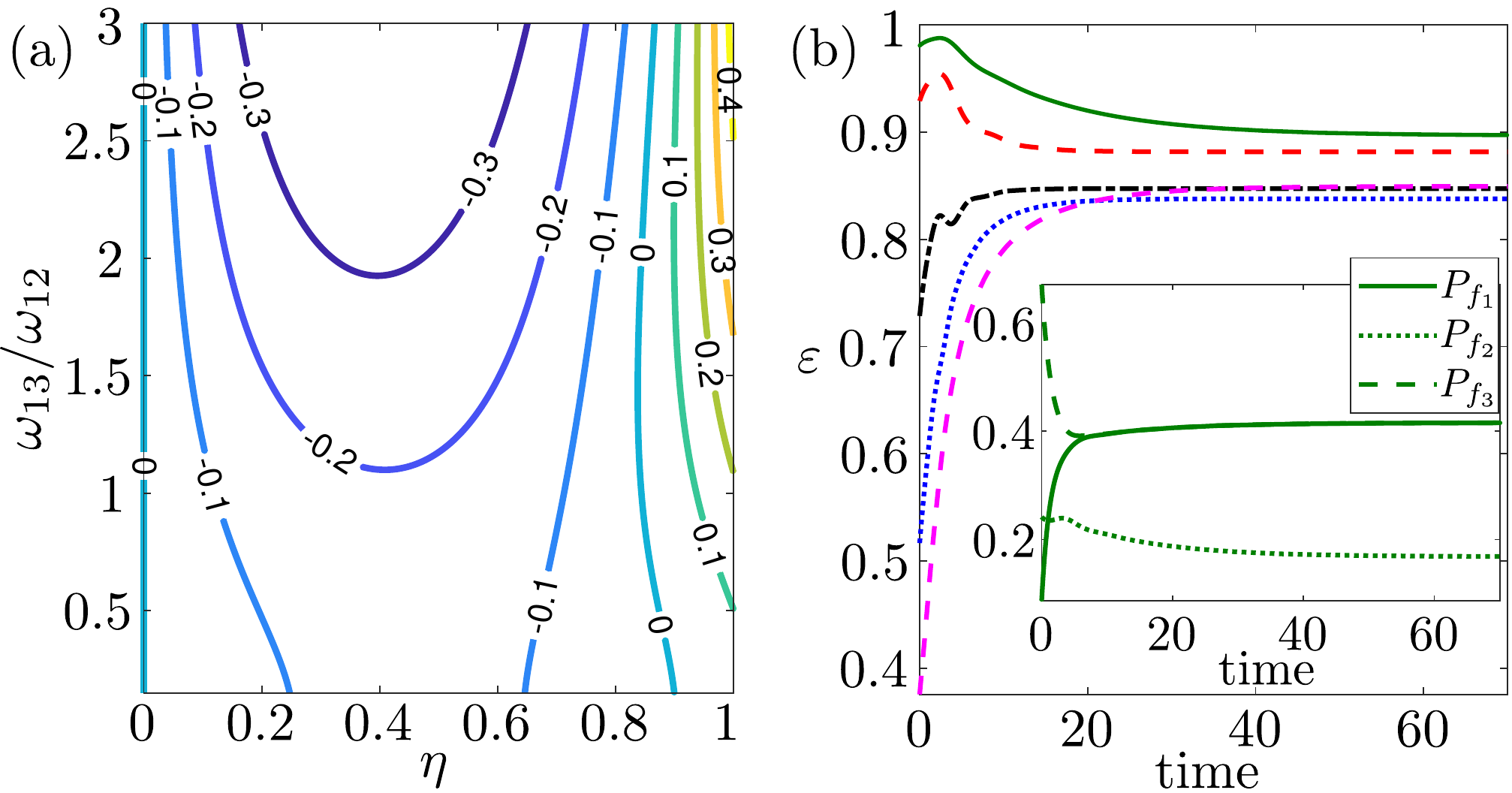}
\caption{Exchange fluctuation relation for a three-level model subjected to spontaneous decay processes. (a) Plot of $g(\eta) \equiv G(\eta)-1$ as a function of $\eta\in[0,1]$ and $\omega_{13}/\omega_{12}\in[0,3]$ at the steady state in the energy basis. (b) $\varepsilon$ as a function of time $t\in[0,80]$ for 4 different values of $\omega_{13}/\omega_{12}$, i.e., $\omega_{13}/\omega_{12}=\{0.21, 0.41, 0.91, 1.51, 2.28\}$, corresponding respectively to the green solid, red dashed, black dash-dotted, blue dotted and magenta dashed lines. Inset: Time behaviour of the probabilities $P_{f_{j}}$ to measure the energy of the system at the final time instant of the TPM protocol, by taking $\omega_{13}=0.21$ and $\omega_{12}=1$. All the parameters used to numerically compute the system dynamics, namely $\omega_{13}$, $\omega_{12}=1$, $\Gamma = 1.5$, $\beta=1$ with $\rho_0 = \exp(-\beta H)/{\rm Tr}[\exp(-\beta H)]$, are in units of $\hbar=1$.
}
\label{fig:scale_factor_SS}
\end{center}
\end{figure}

Under the hypothesis that the system reaches a steady-state in its energy basis, the conditional probabilities $P_{j|i}$ are invariant with respect to the initially measured energy values obtained from the TPM scheme. This means that
\begin{equation}\label{supp_eq_steady}
P_{j|1} = P_{j|2} = P_{j|3} \equiv \widetilde{P}_{j}
\end{equation}
for any $j=1,2,3$, irrespective of the initial state $\rho_0$ and the transient properties of the dissipative dynamics acting on the 3LS.
The validity of Eq.\,(\ref{supp_eq_steady}) implies the following decomposition for the mean exponentiated energy variation $\langle\exp(-\eta\Delta E)\rangle$:
\begin{equation}\label{supp_eq_decomp}
\langle\exp(-\eta\Delta E)\rangle = \left(\sum_{j=1}^{3}\widetilde{P}_{j}e^{-\eta E_j}\right)\left(\sum_{i=1}^{3}P_{i}e^{\eta E_i}\right) \,,
\end{equation}
where $\{E_i\}$ and $\{E_j\}$ are, respectively, the set of measured energy at the initial and final time instants of the TPM scheme, while $P_{i}$ denotes the probability of measuring $E_i$. We have numerically computed the system dynamics with a time duration long enough to ensure that the system has reached a steady-state in the energy basis for each set of system parameters. In Fig.\,\ref{fig:scale_factor_SS}(a), we have plotted $g(\eta) \equiv G(\eta)-1 $ as a function of $\eta$ and $\omega_{13}/\omega_{12}$ at the SSE, by fixing the value of the other parameters (see caption). One can observe that, for each value of $\omega_{13}$ ($\omega_{12}=1$), $g(\eta)$ always has two zeros, where one zero is always $\eta_0 = 0$.
Instead, regarding the other zero of $g(\eta)$, in the numerical simulations it takes a unique time-independent value $\eta=\varepsilon$ that depends only on the initial state $\rho_0$ and the asymptotic steady state.

Finally, to conclude our analysis, we have fixed five different values of $\omega_{13}$ (with $\omega_{12}=1$) and plotted $\varepsilon$ (value of $\eta$ such that $G(\eta)=1$) as a function of time. In Fig.~\ref{fig:scale_factor_SS}(b) one can observe that, as the time increases, $\varepsilon$ becomes a constant (time-independent) value, depending on the specific steady-state reached by the 3LS. The same behaviour occurs also by varying the other dynamics parameters, e.g.\,$\Gamma$ and $\beta$.

\subsection*{Two-level systems: Energy variation distribution and asymptotic effective temperature}
Measuring the conditional probabilities $P_{\uparrow|\uparrow}$ and $P_{\uparrow|\downarrow}$ gives access to the full statistics of $\Delta E$
\begin{equation}\label{eq:probsDeltaE}
P_{\Delta E} \equiv {\rm Prob}(\Delta E) = \sum_{i,j} \delta(\Delta E - \Delta E_{i,j})P_{j|i}P_{i},
\end{equation}
even without directly measuring the output energy for each experiment realization. The latter, indeed, cannot be achieved due to low photon collection efficiency from the NV center.
In the previous equation $P_i$ denotes the probability of measuring $E_i$ at the beginning of the TPM protocol, $P_{j|i}$ is the conditional probability of measuring $E_j$ at the end of the protocol, and the sum is performed over all the possible initial $i$ and final $j$ measured energies.
Once obtained the distribution of the energy variation, we operatively evaluate the mean value of the exponentiated energy fluctuations of the two-level system as
\begin{equation}
\left< e^{-\Delta\beta^{({\rm eff})}\Delta E }\right> = e^{- \Delta\beta^{({\rm eff})}\,\hbar\omega}P_{\hbar\omega} + P_{0} + e^{\Delta\beta^{({\rm eff})}\,\hbar\omega}P_{-\hbar\omega},
\label{eq:LHS}
\end{equation}
where $\Delta\beta^{({\rm eff})}=\beta^{({\rm eff})}_{\rm{in}}-\beta^{({\rm eff})}_\infty$.
For a two-level mixed state $\varrho^{\mathrm{mix}}= p_\uparrow \varrho_\uparrow + (1 - p_\uparrow) \varrho_\downarrow$, an effective inverse temperature is defined as
\begin{equation}
\beta^{({\rm eff})} (p_{\uparrow})= \frac{2}{\hbar\omega}\arctanh(1-2 p_{\uparrow}).
\label{eq:thermalrho}
\end{equation}
This picture of effective temperature does not capture the possible population inversion, and is therefore valid only for the half of the Bloch sphere containing the state at lower energy.
In the experiment, $\beta^{({\rm eff})}_{\rm{in}}$ is defined by the choice of the initial mixed state.
The asymptotic inverse temperature $\beta^{({\rm eff})}_{\infty}$ is extracted from the experimental data after a large enough number of $z$--QPMs, whereby $P_{\uparrow|\uparrow}^\infty= P_{\uparrow|\downarrow}^\infty\equiv P_\uparrow^\infty$, thus ensuring the stationarity of the system final state.
Otherwise, for low photon-absorption probability, the asymptotic temperature can be extracted by finding the initial state for which $\braket{\Delta E}=0$ during the whole evolution.

\bibliography{OQS-Biblio}

\newpage
\section*{Supplemental material}

The effective two-level system that we consider is formed by the $m_S=0$ and $m_S=1$ sublevels of the ground spin state of a Nitrogen-Vacancy (NV) center in diamond. However, the photodynamics of the NV center can be described with a seven-level model~\cite{Manson06}. Here, we characterize the seven-level model and introduce an effective two-level model, appropriate to describe the whole spin dynamics for our experimental protocol, entailing a microwave-driven unitary evolution interspersed with trains of short laser pulses that lead to quantum projective measurements and a dissipation dynamics, as described in the main text.

Then, we show an analytic derivation of the effective final temperature of the out-of-equilibrium steady-state of the NV spin, when applying perfect quantum projective measurements. We also consider a stochastic temporal distribution of projective measurements, as performed in our experiments, and compare the analytic solution with numerical results.


\section{Characterization of the seven-level system}

\begin{figure}
\begin{center}
\includegraphics[scale=0.42]{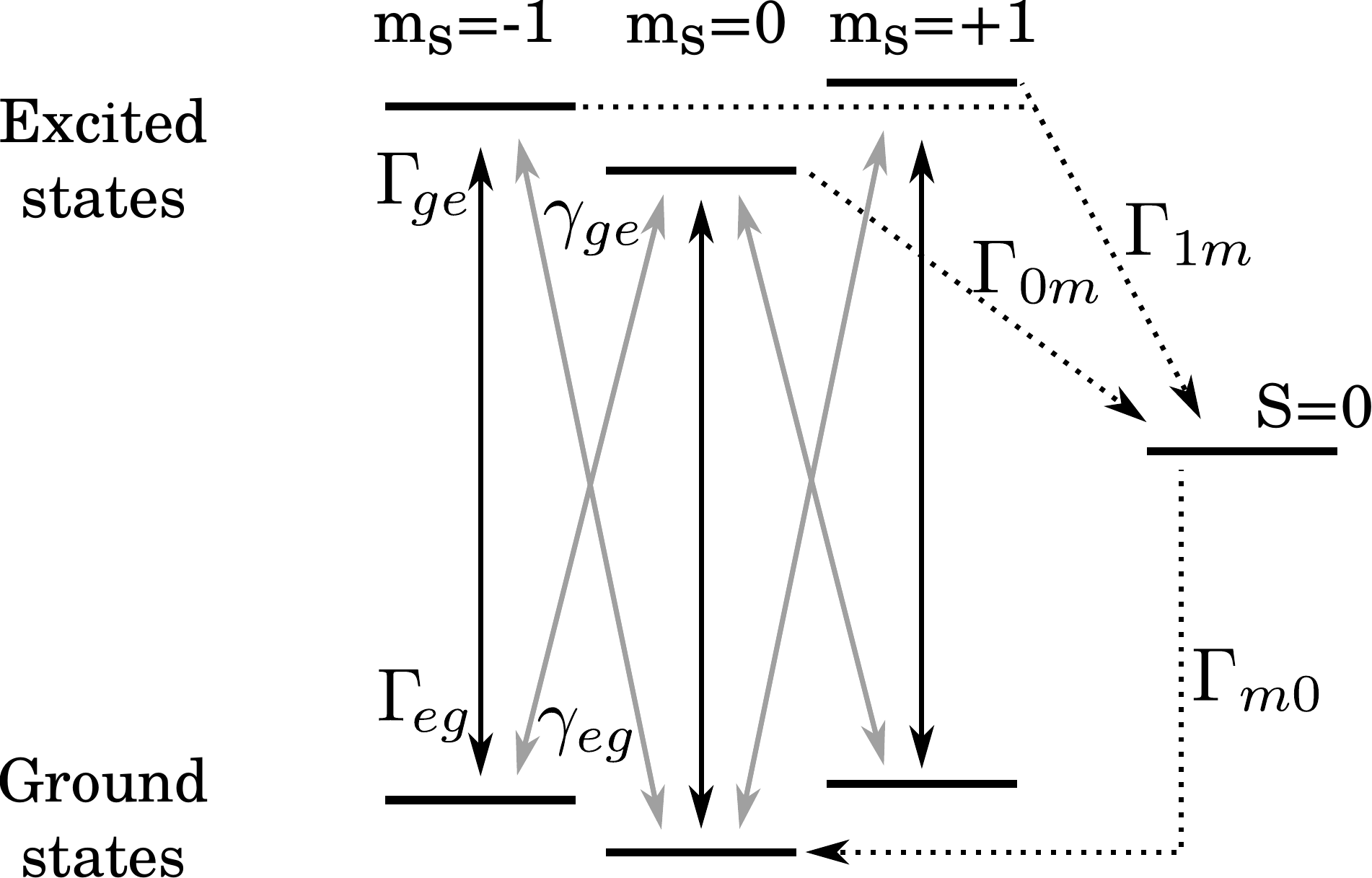}
\caption{Energy levels involved in the excitation and relaxation process during the absorption of green laser light (not to scale). Solid lines represent radiative transitions, while dotted lines represent non-radiative decays. All the solid black arrows are associated to the same decay ({\it resp.,} excitation) rates $\Gamma_{eg}$ ({\it resp.,} $\Gamma_{ge}$), while all the gray arrows are associated to the decay ({\it resp.,} excitation) rates $\gamma_{eg}$ ({\it resp.,} $\gamma_{ge}$).}
\label{fig:energyLevels}
\end{center}
\end{figure}

The electronic seven-level model is composed by the ground-state and excited-state spin triplets, and one metastable spin singlet state, as depicted in Fig.~\ref{fig:energyLevels}. In the following, the order of the levels is $\{g_{+1},g_{0},g_{-1},e_{+1},e_{0},e_{-1},m \}$, where $g$ stands for ground, $e$ for excited, and $m$ for metastable.
The set of parameters that completely describes the system is $X = \{\Gamma_{eg},\Gamma_{1m},\Gamma_{0m},\Gamma_{m0},\theta\}$, together with the excitation rate $\Gamma_{abs}$. $\Gamma_{eg}$ is the spontaneous spin-conserving emission rate and $\Gamma_{ge} = p_{abs} \cdot \Gamma_{eg}$ represents the rate for the electric dipole absorption and stimulated emission, with absorption probability $p_{abs}$. Note that $\Gamma_{abs} \sim \Gamma_{eg}$, the  difference being given by corrections due to spin non-preserving radiative transition
 with rates $\gamma_{eg}$.
Thus, $\tan^2 \theta = \frac{\gamma_{ge}}{\Gamma_{ge}} = \frac{\gamma_{eg}}{\Gamma_{eg}}$ is the relative probability for a spin non-conserving radiative transition to occur. Additional non-radiative transitions involve the spin singlet state, with decay rates $\Gamma_{1m},\Gamma_{0m}$ from $m_S = +1$ and $0$, respectively, towards the metastable, and $\Gamma_{m0}$ from the singlet towards the $m_S = 0$ in the ground state. All the possible transitions are schematically represented in Fig.~\ref{fig:energyLevels}.

The global Hamiltonian describing the radiation-matter coherent interaction for the system can be represented through the matrix:
\begin{equation}
\mathcal{H}_7 = \begin{pmatrix}
    0 & 0 &  0 & \Gamma_{ge} & \gamma_{ge} & 0 & 0\\
    0 & 0 & 0 & \gamma_{ge}/2 & \Gamma_{ge} & \gamma_{ge}/2 & 0 \\
	0 & 0 & 0 & 0 & \gamma_{ge} & \Gamma_{ge} & 0\\
	\Gamma_{ge} & \gamma_{ge}/2 & 0 & 0 & 0 & 0 & 0\\
    \gamma_{ge} & \Gamma_{ge} & \gamma_{ge} & 0 & 0 & 0 & 0\\
	0 & \gamma_{ge}/2 & \Gamma_{ge} & 0 & 0 & 0 & 0\\
	0 & 0 & 0 & 0 & 0 & 0 & 0\\
\end{pmatrix}.
\end{equation}

The spontaneous decays are described by a Lindbladian super-operator, whereby the sum of all the decay routes written in matrix form is:
\begin{equation}
\mathcal{L} = \begin{pmatrix}
    0 & 0 &  0 & \sqrt{\Gamma_{eg}} & \sqrt{\gamma_{eg}/2} & 0 & \sqrt{\Gamma_{m1}}\\
    0 & 0 & 0 & \sqrt{\gamma_{eg}} & \sqrt{\Gamma_{eg}} & \sqrt{\gamma_{eg}} & \sqrt{\Gamma_{m0}} \\
	0 & 0 & 0 & 0 & \sqrt{\gamma_{eg}/2} & \sqrt{\Gamma_{eg}} & \sqrt{\Gamma_{m1}}\\
	0 & 0 & 0 & 0 & 0 & 0 & 0\\
    0 & 0 & 0 & 0 & 0 & 0 & 0\\
	0 & 0 & 0 & 0 & 0 & 0 & 0\\
	0 & 0 & 0 & \sqrt{\Gamma_{1m}} & \sqrt{\Gamma_{0m}} & \sqrt{\Gamma_{1m}} & 0\\
\end{pmatrix}.
\end{equation}
Accordingly, the density matrix evolution follows the differential equation
\begin{equation}\label{supp_diss_dyn}
\frac{d\varrho}{dt}=-i[\mathcal{H}_7,\varrho]+
\Gamma\left(2\mathcal{L}\varrho\mathcal{L}^{\dagger}+\mathcal{L}^{\dagger}\mathcal{L}\varrho + \varrho\mathcal{L}^{\dagger}\mathcal{L}\right),
\end{equation}
with $\hbar$, reduced constant Planck, set to $1$.

\begin{figure*}
\begin{center}
\includegraphics[scale=0.75]{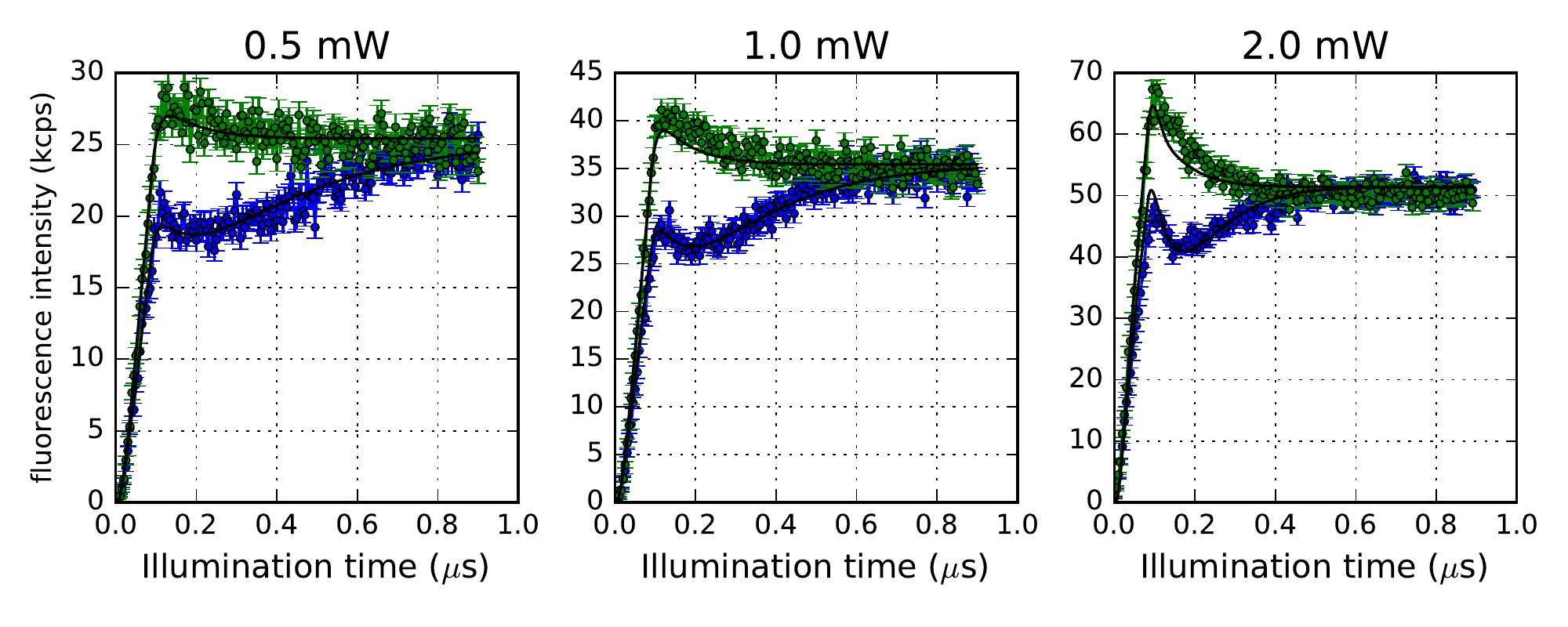}
\caption{Measured PL as a function of the illumination time with green laser light. Experimental data for the $\ket{0}$ (green dots) and $\ket{+1}$ (blue dots) states are shown with simulations with the best fit parameters $X$ and $p_{abs}$ (solid lines).}\label{fig:pl}
\end{center}
\end{figure*}

For the characterization of the decay rates, we performed experiments where we measure the emitted red photo-luminescence (PL) in terms of the illumination time with green laser light.
Results of the experiment are shown in Fig.~\ref{fig:pl}, as well as the simulations after fitting the parameters in $X$, which values are reported in Tab.~\ref{tab:gamma}. The excitation rate depends on the intensity of the green laser light, in particular for the three different experiments shown in Fig.~\ref{fig:pl}, we found that $0.2\Gamma_{ge} \leq\Gamma_{abs}\leq 0.45\Gamma_{ge}$.

\begin{table}
\caption{Parameters resulting from the fit of photo-luminescence with the seven-level model. The decay rates correspond to the transitions represented in Fig.~\ref{fig:energyLevels}.}
\centering
\begin{tabular}{c c}
\hline
\hline
Parameter & Rate\\
\hline
\hline
\vspace{0.2cm}
$\Gamma_{eg}$ & $77$ MHz~\cite{Wolters13}\\
$\Gamma_{1m}$ & $60.4 \pm 0.3 $ MHz\\
$\Gamma_{0m}$ & $9.39 \pm 0.05 $ MHz\\
$\Gamma_{m0}$ & $9.6 \pm 0.05 $ MHz\\
$\theta$ & $0.193 \pm 0.011 $ rad\\
\hline
\hline
\end{tabular}
\par
\label{tab:gamma}
\end{table}

\section{Effective two-level model}
\label{sec:CompBetwModels}

\begin{figure}
\begin{center}
\includegraphics[width=\columnwidth]{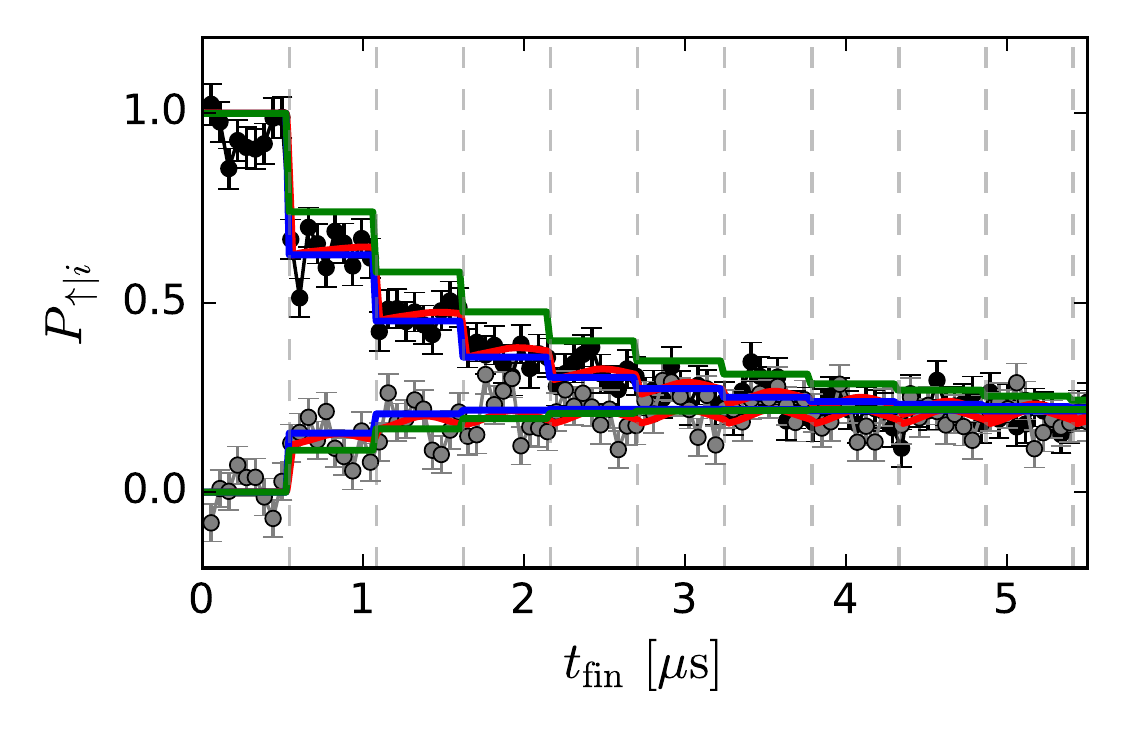}
\caption{Simulations (solid lines) and experimental data (dots), for $\delta=-\Omega$ and inter-pulse time $\tau = 2\pi/\omega$ in the TPM protocol (see main text). Grey dashed lines correspond to the temporal position of the short green laser pulses. The red lines are obtained with the 7-level model, and absorption probability $p_{abs}=0.45$, the green lines with the 2-level model, $p_{abs}=0.45$, the blue lines with the 2-level model and an effective absorption probability equal to $p_{abs}=0.69$.}
\label{fig:comparisonP}
\end{center}
\end{figure}

Once evaluated the quantities responsible of the dynamics of the system, we discuss how they are included in a two-level picture composed by the $m_S = 0, +1$ energy levels of the ground states.
As introduced in the main text, a combined event of spin-preserving absorption and radiative decay corresponds to a quantum projective measurement on the $\sigma_z$ basis ($z$--QPM), characterized by collapse of the wavefunction and coherence cancellation.
The absorption rate is taken into account in the two-level model by considering a finite probability of having a $z$--QPM each time we apply a laser pulse.
On the other hand, the non-radiative decay branch of the seven-level model is included in the two-level model as a dissipation channel, as follows: starting with a pure $m_S = +1$ state, the probability of the measurement outcome to be $\ket{0}$ is $\Gamma_{1m} / (\Gamma_{1m} + \Gamma_{eg}) = 44$~\%. Note that we consider $z$--QPM and dissipation processes to be instantaneous.
Instead, spin non-preserving radiative transitions are taken into account as an additional read-out error, such that when projecting in $m_S = 0$ $(+1)$, the probability to measure the state $m_S = +1$ $(0)$ is $\tan^2 \theta \simeq 3.8 \times 10^{-2}$.
A single trajectory of the two level system can be modeled using a Monte Carlo simulation, where a random number generator is used to select the possible outcomes of the absorption and emission processes, described before, while the unitary part of the evolution is calculated by simply solving the equation of motion. 
The two-level model is able to predict the dynamics of the system by averaging a large enough amount of simulated single trajectories.
In Fig.~\ref{fig:comparisonP} we compare results obtained for the two-point measurement (TPM) protocol with quantum projective measurements ($z$--QPM) both by considering the two-level and the seven-level models.
In particular, in the two-level model an effective (higher) absorption probability is used to include the dynamics of the neglected levels and recover the same results (within the experimental error) to those of the seven-level model, as shown in Fig.~\ref{fig:comparisonP}.

In the limit of absorption probability equal to one, the time distribution of effective consecutive absorptions is no longer stochastic. Hence, it is possible to model the dynamics with an analytic solution that includes $z$--QPMs and dissipation, as will be described in the following section.

\subsection{Non-stochastic limit and analytic solution}
\label{sec:analytical}

\subsubsection{Model for the composition of \texorpdfstring{$z$}{z}--QPMs and unitary evolutions}

The protocol performed in the experimental runs is based on applying $z$--QPMs alternated with evolutions under the Hamiltonian $\mathcal{H}=\omega(\cos\alpha\sigma_{z}-\sin\alpha\sigma_{x})$. The quantum system evolves under the composition of unitary dynamics and measurement processes, which on average can be modelled by Lindbladian jump operators. As we measure $\sigma_z$, the two measurement projectors are $\ket{0}\!\!\bra{0}$ and $\ket{1}\!\!\bra{1}$ and thus the jump operators are just equal to $\sqrt{\Gamma}\ket{0}\!\!\bra{0}$ and $\sqrt{\Gamma}\ket{1}\!\!\bra{1}$ acting for the time $t_{L}$. With this formalism, ideal projective measurements are obtained in the limit $\Gamma\,t_{L}\rightarrow\infty$, giving the super-operator
\begin{equation}\label{superop_M}
\widetilde{M} \equiv \frac{1}{2}(I_{4}+\sigma_{z}\otimes\sigma_{z}),
\end{equation}
with $I_{4}$ denoting the identity matrix in the $4\times4$ space.
We have here assumed to adopt super-operators that act not on the density operator $\rho$ of the two-level system under investigation but on its vectorization. This choice justifies the need to define the measurement projector $\widetilde{M}$ (indeed, $\widetilde{M}^2 = \widetilde{M}$) in an Hilbert space with dimension $4$. In this mathematical space (identified by the $\widetilde{(\cdot)}$ above the symbols), the unitary evolution of the system is thus equal to $\widetilde{U} \equiv e^{-i\widetilde{\mathcal{H}}t}$ ($\hbar$ set to 1), with
\begin{equation}
\widetilde{\mathcal{H}} \equiv \mathcal{H}\otimes I_{2} - I_{2}\otimes\mathcal{H}^{\ast},
\end{equation}
where $(\cdot)^{\ast}$ denotes complex conjugation. As a result, after have prepared the system in a thermal state at temperature $\beta_{\rm in}$, the super-operator governing its evolution is given by the application for $n$ times, with $n$ number of measurements, of the composition $\widetilde{U}\widetilde{M}$, allowing us to introduce the super-operator $\widetilde{S} \equiv (\widetilde{U}\widetilde{M})^{n}$. After the evolution, we measure the conditional probability
\begin{equation}
P_{\uparrow|\uparrow} = \langle\bra{\uparrow}\widetilde{S}\ket{\uparrow}\rangle
\end{equation}
for the NV-center to be in the energy eigenstate $\ket{\uparrow}\rangle$ by starting from $\ket{\uparrow}\rangle$, where $\ket{\cdot}\rangle$ denotes the vectorization of $(\cdot)$, and the energy eigenstates $\ket{\uparrow}\rangle$ and $\ket{\downarrow}\rangle$ can be written as
\begin{equation}
\ket{\uparrow}\rangle = \frac{1}{2}[1+\cos(\alpha),-\sin(\alpha),-\sin(\alpha),1-\cos(\alpha)]
\end{equation}
and
\begin{equation}
\ket{\downarrow}\rangle = \frac{1}{2}[1-\cos(\alpha),+\sin(\alpha),+\sin(\alpha),1+\cos(\alpha)].
\end{equation}
From $P_{\uparrow|i}$, since we assume a two-level system, we can compute the full statistics of energy variation $\Delta E$, as well as the QJE $\langle e^{-\beta_{\rm in}\Delta E}\rangle$. Here, it is worth noting that, being $\widetilde{M}$ a projector, $\widetilde{S}=\widetilde{M}(\widetilde{U}\widetilde{M})^{n}=(\widetilde{M}\widetilde{U}\widetilde{M})^{n}$. Therefore, by introducing for calculation purposes the quantities $\mu \equiv 1-2\sin^{2}(\alpha)\sin^{2}(\frac{\omega \tau}{2})$ and $N \equiv \frac{1}{2}(\sigma_{x}\otimes\sigma_{x} - \sigma_{y}\otimes\sigma_{y})$, we can derive an analytical expression for $\widetilde{S}$, i.e.
\begin{equation}
\widetilde{S} = (\widetilde{M}\widetilde{U}\widetilde{M})^{n} = \frac{1}{2}\left[(1+\mu^{n})\widetilde{M}+(1-\mu^{n})N\right].
\end{equation}

\subsubsection{Model for the composition of \texorpdfstring{$z$}{z}--QPMs and unitary evolutions with dissipation}

We now introduce the additional dissipative channel associated to each absorbed laser pulse.
The consequent decay in the $\ket{0}$ state can be modeled on average by the Lindbladian jump operator $\sqrt{\Gamma_{\mathcal{D}}}\ket{0}\!\!\bra{1}$ (within the space of the two-level system) acting during the same laser time. This gives the super-operator $\tilde{\mathcal{D}}$ (defined in the Hilbert space with dimension $4$) allowing us to introduce a different measurement super-operator, i.e.\,$\widetilde{M}_{\mathcal{D}} \equiv \widetilde{\mathcal{D}}\widetilde{M} = \widetilde{M}\widetilde{\mathcal{D}}$, able to model the measurement process with dissipation. Therefore, the expression of the super-operator for the global dynamical evolution of the NV-center is equal to
\begin{equation}\label{S_diss}
\widetilde{S}_{\mathcal{D}} = (\widetilde{U}\widetilde{M}_{\mathcal{D}})^{n} = \widetilde{M}\widetilde{\mathcal{D}}^{1/2}
(\widetilde{M}\widetilde{\mathcal{D}}^{1/2}\widetilde{U}\widetilde{\mathcal{D}}^{1/2}\widetilde{M})^{n-1}\widetilde{\mathcal{D}}^{1/2}\widetilde{M},
\end{equation}
where in the r.h.s.\,of Eq.\,(\ref{S_diss}) we have not considered the last unitary evolution, since at the end of the protocol an energy measurement is applied. Moreover, by denoting with $i$ and $j$ the indices for the states of the system at the initial and final energy measurement of the protocol,
the conditional probability $P_{j|i}$ can be written as
\begin{equation}
P_{j|i} = \frac{1}{2}[1 + z_{j}\cos\alpha(R_{n} + z_{i}\mu^{n-1}e^{-n\Gamma_{\mathcal{D}}t_{L}}\cos\alpha)],
\end{equation}
where
\begin{equation}
R_{n} \equiv \frac{1-e^{-\Gamma_{\mathcal{D}}t_{L}}}{1-\mu e^{-\Gamma_{\mathcal{D}}t_{L}}}(1-\mu^{n}e^{-n\Gamma_{\mathcal{D}}t_{L}})
\end{equation}
and $z_{j}$, $z_{i}$ correspond to $+$ or $-$ for $\ket{j}\rangle$ and $\ket{i}\rangle$ equal to $\ket{\uparrow}\rangle$ or $\ket{\downarrow}\rangle$. In conclusion, in the limit of $n\rightarrow\infty$, we  recover Eq.\,(4) of the main text.
\begin{figure*}
\begin{center}
\includegraphics[width=0.72\textwidth]{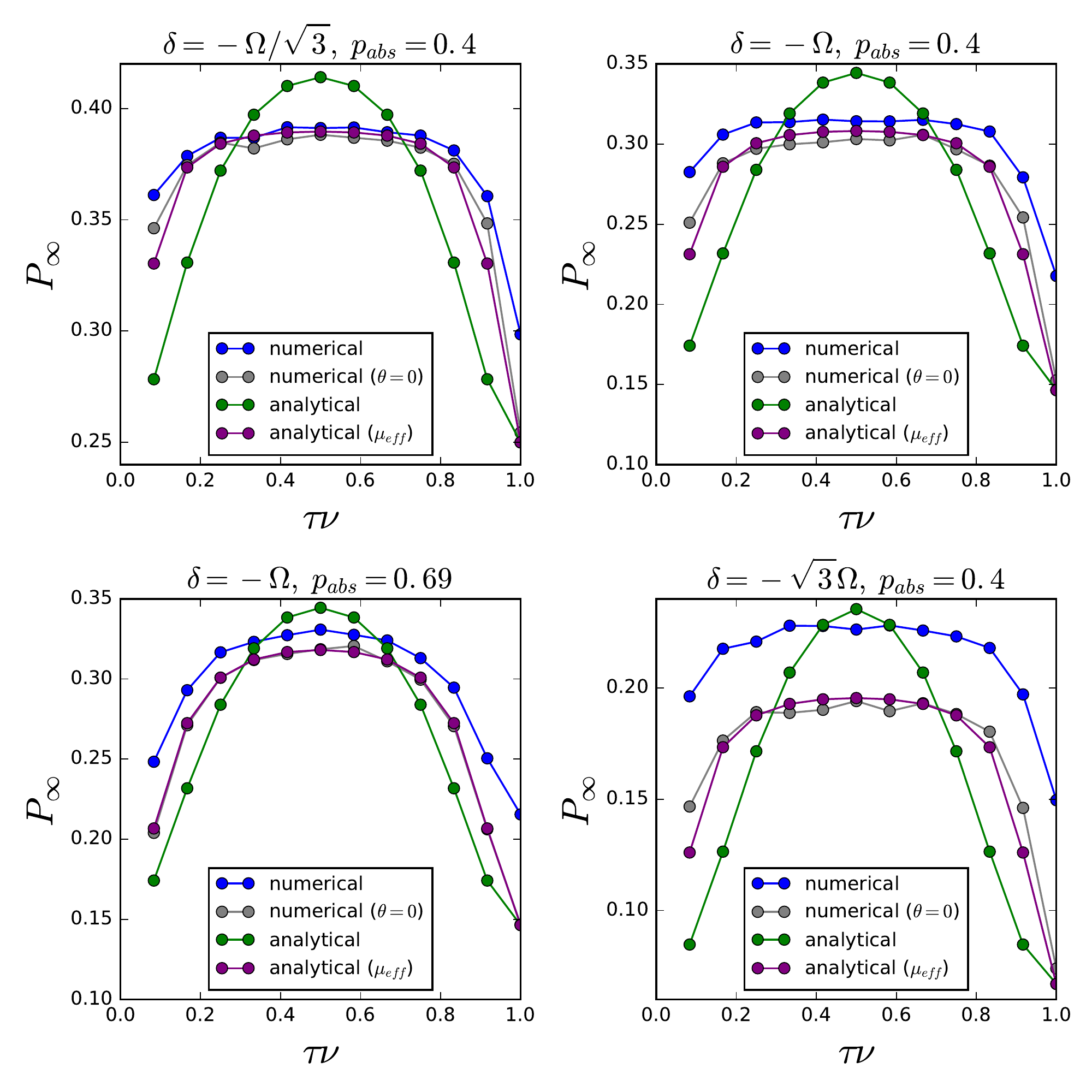}
\caption{Comparison between numerical and analytical calculation of the asymptotic probability for the spin to be in the $\ket{\uparrow}$ state, as a function of the time between laser pulses $\tau$.  Respectively in blue and gray: Numerical simulation with the complete model, and without non-preserving spin transitions, i.e.\,$\theta=0$ (see Sec.~\ref{sec:CompBetwModels}). In green and purple: Analytical solution for perfect absorption, and finite absorption ($\mu_{eff}(p_{abs})$, see text), respectively.
Time $\tau$ is scaled in terms of energy $\nu=\omega/2\pi$ ($\hbar=1$).}
\label{fig:meanMu}
\end{center}
\end{figure*}

\subsection{Stochastic limit}

The finite absorption probability can be accounted in the analytical solution by considering a mean time-interval of free evolution between effective $z$--QPMs, instead of the fixed one used in Sec~\ref{sec:analytical}.
The probability of having $k$ effective absorptions for $n$ laser pulses, each with an absorption probability $p_{abs}$, is obtained from a binomial distribution. From this information we also extract the probability distribution $f_\ell(n,p_{abs})$, that provides the probabilities of having $\ell$ consecutive time intervals $\tau$ between laser absorptions.
We then obtain the effective value of $\mu$ as:
\begin{equation}
\mu_{eff}(p_{abs}) = \sum_{\ell=1}^n \mu_i \, f_\ell(n,p_{abs})
\end{equation}
where $\mu_\ell = 1-2\sin^{2}(\alpha)\sin^{2}(\frac{\ell \tau\omega }{2})$ and $\sum_{\ell=1}^n f_\ell(n,p_{abs})=1$.
In Fig.~\ref{fig:meanMu} we present the comparison between the numerical simulation and the analytical solution using both $\mu_{eff}(p_{abs})$ (finite absorption probability) and $\mu$ (perfect absorption).
This effective $\mu$ correctly takes into account the finite absorption probability, however the analytical solution is still missing a description for the non-preserving spin radiative transition process, which is more important as we reduce the value of $\alpha$, as shown in Fig.~\ref{fig:meanMu}.

\end{document}